\documentclass[10pt,journal,final,twocolumn]{IEEEtran}

\usepackage{graphics}
\usepackage{subfigure}
\usepackage{epstopdf}
\usepackage{epsfig}
\usepackage[cmex10]{amsmath}
\usepackage{amssymb}
\usepackage{times}
\usepackage{ntheorem}
\usepackage{pifont}
\usepackage{stfloats}
\usepackage{bm}
\usepackage{color}
\usepackage{makecell}
\usepackage{array}
\usepackage{multirow}

\begin{document}
%
\title{\huge Deep Multi-Stage CSI Acquisition for Reconfigurable Intelligent Surface Aided MIMO Systems}

\author{\IEEEauthorblockN{
Shen~Gao, Peihao~Dong,~\IEEEmembership{Member,~IEEE}, Zhiwen~Pan,~\IEEEmembership{Member,~IEEE},
and Geoffrey Ye Li,~\IEEEmembership{Fellow,~IEEE}
}

\vspace{-0.5cm}
\thanks{

S. Gao and Z. Pan are with the National Mobile Communications Research Laboratory, Southeast University, Nanjing 210096, China, and also with the Purple Mountain Laboratories, Nanjing 211100, China (e-mail: gaoshen@seu.edu.cn; pzw@seu.edu.cn).

P. Dong is with the  College of Electronic and Information Engineering, Nanjing University of Aeronautics and Astronautics, Nanjing 211106, China (e-mail: phdong@nuaa.edu.cn).

G. Y. Li is with Department of Electrical and Electronic Engineering, Imperial College London, London SW7 2AZ, U.K. (e-mail: Geoffrey.Li@imperial.ac.uk).
}
\vspace{-0.3cm}
}

\IEEEtitleabstractindextext{%
\begin{abstract}
This article aims to reduce huge pilot overhead when estimating the reconfigurable intelligent surface (RIS) relayed wireless channel. Motivated by the compelling grasp of deep learning in tackling nonlinear mapping problems, the proposed approach only activates a part of RIS elements and utilizes the corresponding cascaded channel estimate to predict another part. Through a synthetic deep neural network (DNN), the direct channel and active cascaded channel are first estimated sequentially, followed by the channel prediction for the inactive RIS elements. A three-stage training strategy is developed for this synthetic DNN. From simulation results, the proposed deep learning based approach is effective in reducing the pilot overhead and guaranteeing the reliable estimation accuracy.
\end{abstract}

}

\maketitle

\IEEEdisplaynontitleabstractindextext

%
\IEEEpeerreviewmaketitle

\vspace{-0.2cm}
\section{Introduction}

\IEEEPARstart{T}{he} recent advent of reconfigurable intelligent surface (RIS) has stirred up a plethora of research activities since it has potential to boost the network performance and reduce the cost \cite{Q. Wu}--\cite{C. Huang_b}. The general RIS consists of an inexpensive smart surface, usually made of either tiny antenna elements or metamaterials, and some low power circuits. Through real-time reflecting adaption by an external controller, RIS bears the ability of manipulating the phase and amplitude of the impinging signal in order to focus the signal energy at the receiver as well as to mitigate the interference and security threats \cite{M. Di Renzo}--\cite{Z. Yang}.

The above-mentioned brightening advantages of RIS build on the reliable channel state information (CSI). However, the CSI acquisition is challenging since the RIS-relayed link is a cascaded channel whose structure differs from the single-hop channel, which makes most channel estimation approaches exploiting single-hop channel statistics malfunction. In addition, the huge pilot overhead incurred by the RIS is a vital limit thwarting the dramatic performance improvement. In \cite{D. Mishra}, a simple on/off operation mode for the RIS is proposed to sequentially estimate the cascaded channel associated with each active RIS element. By exploiting the array gain of the RIS, an efficient least-square (LS) based approach is developed in \cite{B. Zheng} and \cite{T. L. Jensen}, which adapts the reflection coefficient of each RIS element as per a discrete Fourier transformation (DFT) matrix to improve the estimation accuracy. In \cite{A. Taha}, compressed sensing (CS) is exploited to separately estimate the sparse channel of each hop by endowing the RIS with signal processing ability. In \cite{P. Wang}, a novel sparse form of the cascaded channel is uncovered to enable the sparse channel recovery with the reduced pilot overhead provided that the sparsity of each hop is known.

As one of the key technologies underlaying the pathway to smart radio, deep learning (DL) based signal processing has sparked a revolution in wireless communications \cite{H. Ye}--\cite{S. Gao}, and thus inspires some attempts to address the channel estimation problem for RIS-aided systems. In \cite{A. M. Elbir} and \cite{N. K. Kundu}, deep convolutional neural network (CNN) is applied to refine the LS channel estimate to further improve the accuracy. A complex-valued denoising CNN is proposed in \cite{S. Liu} to enhance the CS-based estimate for the broadband user-RIS channel. 

It can be seen that the CS-based channel estimation approaches highly rely on channel statistics as the prior knowledge and will suffer from performance degradation in the practical complicated scenarios. Although the LS-based approaches get rid of this dependence, they still have not well traded off between the estimation accuracy and the pilot overhead. The success of DL in estimating the single-hop channels \cite{P. Dong}, \cite{S. Gao} has implied its potential to address the above-mentioned problems and thus enlightens us to figure out a solution along with this line. The main novelty and contribution of this article can be summarized as follows:
\vspace{-0.1cm}
\begin{itemize}[\IEEEsetlabelwidth{Z}]
\item[1)] With partially activated RIS elements to reduce the pilot overhead, we propose a three-stage CSI acquisition framework successively including estimation of the direct channel, estimation of the cascaded channel for active RIS elements, and prediction of the cascaded channel for inactive RIS elements, and develop a synthetic deep neural network (DNN) to realize it.

\item[2)] For the prediction stage, we discover a simple yet efficient mapping relationship from the perspective of each base station (BS) antenna, which facilitates the DNN training under the accumulative estimation errors propagated from the previous stages and thus achieves the superior estimation accuracy.
\end{itemize}
\vspace{-0.1cm}

\emph{Notations}: In this article, we use upper and lower case boldface letters to denote matrices and vectors, respectively. $(\cdot)^T$, $(\cdot)^H$, $\|\cdot\|_{F}$, and $\mathbb{E}\{\cdot\}$ represent the transpose, conjugate transpose, Frobenius norm, and expectation, respectively. $\|\cdot\|$ denotes the Euclidian norm of a vector. $\textrm{diag}(\mathbf{x})$ transforms vector $\mathbf{x}$ to a diagonal matrix. $\mathbf{0}_{N}$ denotes an all-zero column vector with $N$ entries. $\mathbf{1}_{N}$ denotes a column vector with each of $N$ entries equal to $1$. $\mathbf{I}_{N}$ denotes an $N\times N$ identity matrix. $\mathcal{CN}(0,\sigma^2)$ represents a circular symmetric complex Gaussian distribution with variance $\sigma^2$. $|\mathcal{X}|$ denotes the cardinality of set $\mathcal{X}$.

\vspace{-0.2cm}
\section{System Model}

\begin{figure}[t]
\centering
\includegraphics[width=2.3in]{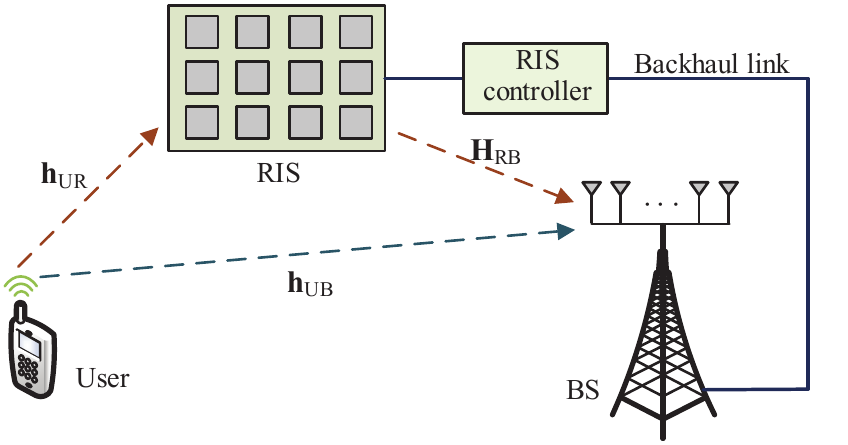}
\caption{A RIS-aided MIMO uplink system.}\label{System_model}
\vspace{-0.4cm}
\end{figure}

We consider a RIS-aided multiple-input multiple-output (MIMO) uplink system shown in Fig.~\ref{System_model}, where a single-antenna user transmits signals to a BS with $M$ antennas and a RIS with $N$ reflecting elements is deployed to enhance the communication quality by using extremely low power and cost. The reflecting coefficient of each RIS element can be adapted by the BS via a controller based on the real-time CSI in order to smartly rotate the phase of the incident signal. In Fig.~\ref{System_model}, the MIMO system is operated in the time division duplex mode, where the BS acquires the CSI resorting to the pilot signals transmitted by the user.

At the $i$th time instant of a coherence interval, the received pilot signal at the BS is expressed as
\vspace{-0.15cm}
\begin{eqnarray}
\label{eqn_yi}
\mathbf{y}_{i}=\sqrt{P}(\mathbf{h}_{\textrm{UB}}+\mathbf{H}_{\textrm{RB}}\textrm{diag}(\boldsymbol{\phi}_{i})\mathbf{h}_{\textrm{UR}})x_{i}+\mathbf{z}_{i},
\end{eqnarray}
where $P$, $x_{i}$, $\boldsymbol{\phi}_{i}$, and $\mathbf{z}_{i}\sim \mathcal{CN}(0, \sigma_{0}^2\mathbf{I}_{M})$ denote the user transmit power, the transmitted pilot signal, the reflecting coefficient vector of the RIS, and the additive white Gaussian noise (AWGN) at the BS, respectively. From \cite{B. Zheng}, the $n$th entry of $\boldsymbol{\phi}_{i}\in \mathbb{C}^{N\times 1}$ corresponds to the reflecting coefficient of the $n$th RIS element and is written as $\phi_{ni}=\beta_{ni}e^{j\varphi_{ni}}$ with $\beta_{ni}\in [0,1]$ and $\varphi_{ni}\in (0,2\pi]$ accounting for the reflecting amplitude and phase rotation, respectively. In this article, we set $\beta_{ni}=1$ for the active RIS elements to avoid the energy loss and simplify the RIS hardware structure. $\mathbf{h}_{\textrm{UB}}\in \mathbb{C}^{M\times 1}$, $\mathbf{h}_{\textrm{UR}}\in \mathbb{C}^{N\times 1}$, and $\mathbf{H}_{\textrm{RB}}\in \mathbb{C}^{M\times N}$ denote the channel from the user to the BS, the channel from the user to the RIS, and the channel from the RIS to the BS, respectively.

To facilitate further signal processing, $\mathbf{y}_{i}$ is rewritten as
\vspace{-0.15cm}
\setlength{\arraycolsep}{0.1em}
\begin{eqnarray}
\label{eqn_yi2}
\mathbf{y}_{i}&&=\sqrt{P}(\mathbf{h}_{\textrm{UB}}+\mathbf{H}_{\textrm{RB}}\textrm{diag}(\mathbf{h}_{\textrm{UR}})\boldsymbol{\phi}_{i})x_{i}+\mathbf{z}_{i}\nonumber\\
&&\triangleq\sqrt{P}(\mathbf{h}_{\textrm{UB}}+\mathbf{G}\boldsymbol{\phi}_{i})x_{i}+\mathbf{z}_{i}.
\end{eqnarray}
Note that it is more tractable to estimate the equivalent cascaded channel $\mathbf{G}$ instead of estimating $\mathbf{H}_{\textrm{RB}}$ and $\mathbf{h}_{\textrm{UR}}$ separately and $\mathbf{G}$ can be directly used for the beamforming design during the data transmission. Hence, we will focus on the acquisition of $\mathbf{h}_{\textrm{UB}}$ and $\mathbf{G}$ hereinafter.

Then the received pilot signals at the BS during $\tau+1$ time instants are given in matrix form as
\vspace{-0.15cm}
\begin{eqnarray}
\label{eqn_Y}
\mathbf{Y}=\sqrt{P}(\mathbf{h}_{\textrm{UB}}\mathbf{1}_{\tau+1}^{T}+\mathbf{G}\boldsymbol{\Psi})\mathbf{X}+\mathbf{Z},
\end{eqnarray}
where $\boldsymbol{\Psi}=[\boldsymbol{\phi}_{1},\ldots,\boldsymbol{\phi}_{\tau+1}]$ with $\boldsymbol{\phi}_{1}=\mathbf{0}_{N}$, $\mathbf{X}=\textrm{diag}([x_{1},\ldots,x_{\tau+1}])$, and $\mathbf{Z}=[\mathbf{z}_{1},\ldots,\mathbf{z}_{\tau+1}]$. Without loss of generality, we assume $\mathbf{X}=\mathbf{I}_{\tau+1}$ for simplicity, which yields
\vspace{-0.15cm}
\begin{eqnarray}
\label{eqn_Y_simp}
\mathbf{Y}=\sqrt{P}(\mathbf{h}_{\textrm{UB}}\mathbf{1}_{\tau+1}^{T}+\mathbf{G}\boldsymbol{\Psi})+\mathbf{Z}.
\end{eqnarray}

Based on the general pilot transmission model sketched above, we investigate how to acquire reliable $\mathbf{h}_{\textrm{UB}}$ and $\mathbf{G}$ with low pilot overhead via DL in the following.

\vspace{-0.2cm}
\section{DL-Based Three-Stage CSI Acquisition}

In this section, a novel three-stage CSI acquisition framework is proposed based on DL. We will first shed light on the basic idea of the framework and then design a synthetic DNN as its backbone for CSI acquisition. Finally, the online testing procedure of the framework is described.

\vspace{-0.2cm}
\subsection{Basic Idea}

In the proposed framework, $\mathbf{h}_{\textrm{UB}}$ is first estimated via a DNN while all RIS elements are turned off. After then, a part of RIS elements are activated with their reflection coefficients adapted as per the DFT matrix \cite{B. Zheng}, \cite{T. L. Jensen} so that the BS estimates the corresponding equivalent cascaded channel by using another DNN\footnote{By turning on only a part of RIS elements, relatively accurate LS channel estimate for these active RIS elements can be obtained even with reduced pilot overhead, which can be further exploited by the elaborated DNNs to construct the complete channel.}. Finally, the DNN-based channel prediction is conducted to retrieve the equivalent cascaded channel associated with those inactive RIS elements. These three DNNs hook up in a sequential manner and compose a synthetic DNN. To achieve a good overall estimation accuracy, we design and train the DNNs separately in three stages\footnote{The DNNs are trained offline and thus the computational cost of training is relatively trivial.}.

\vspace{-0.2cm}
\subsection{Three-Stage DNN Design}

\begin{figure*}[t]
\centering
\includegraphics[width=6.2in]{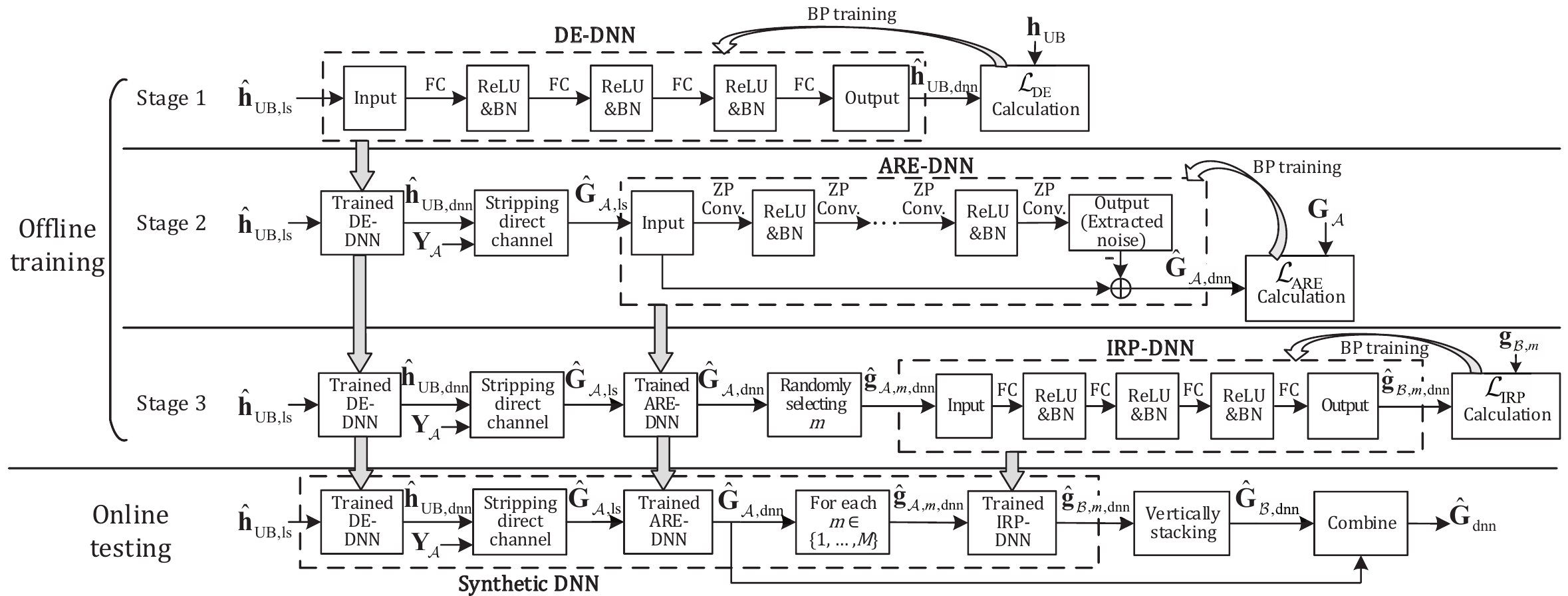}
\caption{DL-based three-stage CSI acquisition framework.}\label{CSI_framework}
\vspace{-0.4cm}
\end{figure*}

At the beginning of this subsection, we outright present the overall structure of the proposed CSI acquisition framework in Fig.~\ref{CSI_framework} to facilitate the elaboration on the synthetic DNN design in each stage.

\emph{1) Stage 1:} At the first time instant during pilot transmission, all RIS elements are turned off so that the BS can estimate $\mathbf{h}_{\textrm{UB}}$ from the received pilot signal, which is the first column of $\mathbf{Y}$ in (\ref{eqn_Y_simp}) and is written as
\vspace{-0.15cm}
\begin{eqnarray}
\label{eqn_y1}
\mathbf{y}_{1}=\sqrt{P}\mathbf{h}_{\textrm{UB}}+\mathbf{z}_{1}.
\end{eqnarray}
\vspace{-0.1cm}
Then the LS estimate of $\mathbf{h}_{\textrm{UB}}$ can be obtained as $\hat{\mathbf{h}}_{\textrm{UB,ls}}=\frac{\mathbf{y}_{1}}{\sqrt{P}}$.

As shown in Fig.~\ref{CSI_framework}, $\hat{\mathbf{h}}_{\textrm{UB,ls}}$ is then input into a \textbf{D}irect Channel \textbf{E}stimation DNN (DE-DNN) in attempts to approximate the true channel $\mathbf{h}_{\textrm{UB}}$. Thus DE-DNN is trained with the sample tuple, $\langle\hat{\mathbf{h}}_{\textrm{UB,ls}}, \mathbf{h}_{\textrm{UB}}\rangle$, and the loss function,
\vspace{-0.15cm}
\begin{eqnarray}
\label{eqn_mse_DE}
\mathcal{L}_{\textrm{DE}}=\frac{1}{N_{\textrm{tr}}}\sum_{n=1}^{N_{\textrm{tr}}}\|\mathbf{h}_{\textrm{UB}}^{(n)}-\hat{\mathbf{h}}_{\textrm{UB,dnn}}^{(n)}\|^2,
\end{eqnarray}
where $N_{\textrm{tr}}$ denotes the number of training samples, $\hat{\mathbf{h}}_{\textrm{UB,dnn}}^{(n)}$ denotes the channel approximated by DE-DNN, and the superscript $(n)$ indicates the $n$th sample. In more detail, DE-DNN is fully-connected (FC) and includes three hidden layers, each of which applies rectified linear unit (ReLU) activation function and batch normalization (BN) to avoid gradient vanishing and overfitting. The output layer does not apply any activation function so that the label values need not to be tailored to fit the activation function.

\emph{2) Stage 2:} In this stage, only a part of RIS elements, denoted by set $\mathcal{A}$ with $N_{1}=|\mathcal{A}|$, are activated to reflect the pilot signals transmitted by the user to the BS. Considering the efficient DFT matrix based reflection mode for the active elements \cite{B. Zheng}, \cite{T. L. Jensen}, the received pilot signals at the BS from the second to the $(N_{1}+1)$th time instant can be expressed as
\vspace{-0.15cm}
\begin{eqnarray}
\label{eqn_Y_act}
\mathbf{Y}_{\mathcal{A}}=\sqrt{P}(\mathbf{h}_{\textrm{UB}}\mathbf{1}_{N_{1}}^{T}+\mathbf{G}_{\mathcal{A}}\boldsymbol{\Phi}_{N_{1}})+\mathbf{Z}_{\mathcal{A}},
\end{eqnarray}
where $\mathbf{G}_{\mathcal{A}}\in \mathbb{C}^{M\times N_{1}}$ denotes the equivalent cascaded channel associated with the active RIS elements whose columns are fetched from $\mathbf{G}$ according to the indexes in $\mathcal{A}$, the DFT matrix, $\boldsymbol{\Phi}_{N_{1}}\in \mathbb{C}^{N_{1}\times N_{1}}$, indicates the reflection coefficients, and $\mathbf{Z}_{\mathcal{A}}$ is the corresponding AWGN.

As $\mathbf{h}_{\textrm{UB}}$ has been estimated by the DE-DNN, its estimate $\hat{\mathbf{h}}_{\textrm{UB,dnn}}$ will be subtracted from $\mathbf{Y}_{\mathcal{A}}$, which yields the LS estimate of $\mathbf{G}_{\mathcal{A}}$, i.e.,
\vspace{-0.15cm}
\begin{eqnarray}
\label{eqn_GA_ls}
\hat{\mathbf{G}}_{\mathcal{A},\textrm{ls}}&&=\left(\frac{\mathbf{Y}_{\mathcal{A}}}{\sqrt{P}}-\hat{\mathbf{h}}_{\textrm{UB,dnn}}\mathbf{1}_{N_{1}}^{T}\right) \frac{\boldsymbol{\Phi}_{N_{1}}^{H}}{N_{1}}\nonumber\\
&&=\mathbf{G}_{\mathcal{A}}+\left((\mathbf{h}_{\textrm{UB}}-\hat{\mathbf{h}}_{\textrm{UB,dnn}})\mathbf{1}_{N_{1}}^{T} +\frac{\mathbf{Z}_{\mathcal{A}}}{\sqrt{P}}\right)\frac{\boldsymbol{\Phi}_{N_{1}}^{H}}{N_{1}}.
\end{eqnarray}
From Fig.~\ref{CSI_framework}, $\hat{\mathbf{G}}_{\mathcal{A},\textrm{ls}}$ will be refined by an \textbf{A}ctive \textbf{R}IS Channel \textbf{E}stimation DNN (ARE-DNN) to output a version closer to the true channel, $\mathbf{G}_{\mathcal{A}}$. Similarly, the loss function for ARE-DNN training is given by
\vspace{-0.15cm}
\begin{eqnarray}
\label{eqn_mse_ARE}
\mathcal{L}_{\textrm{ARE}}=\frac{1}{N_{\textrm{tr}}}\sum_{n=1}^{N_{\textrm{tr}}}\|\mathbf{G}_{\mathcal{A}}^{(n)}-\hat{\mathbf{G}}_{\mathcal{A},\textrm{dnn}}^{(n)}\|_{F}^2,
\end{eqnarray}
where $\hat{\mathbf{G}}_{\mathcal{A},\textrm{dnn}}^{(n)}$ denotes the output of ARE-DNN.

Since $\hat{\mathbf{G}}_{\mathcal{A},\textrm{ls}}$ is corrupted by the residual estimation error of Stage 1 in addition to AWGN, we invoke the more efficient residual network structure to design ARE-DNN. As shown in Fig.~\ref{CSI_framework}, the input, $\hat{\mathbf{G}}_{\mathcal{A},\textrm{ls}}$, flows in ARE-DNN through two ways, one of which includes several convolutional layers to successively distill the aggregated noise from $\hat{\mathbf{G}}_{\mathcal{A},\textrm{ls}}$ while another one is a shortcut representing the identity mapping. The two ways intersect by subtracting the extracted noise from the input, $\hat{\mathbf{G}}_{\mathcal{A},\textrm{ls}}$, and then a more purified channel, $\hat{\mathbf{G}}_{\mathcal{A},\textrm{dnn}}$, can be obtained. Specifically, the first way consists of eight zero padding (ZP) convolutional layers for feature extraction \cite{P. Dong}. Each of the first seven layers applies $64$ $3\times 3$ kernels, ReLU activation function and BN while the last layer only applies $2$ $3\times 3$ kernels and directly output the filtered results. Simulation trails show that further increasing the number of convolutional layers will not improve the performance. In addition, using the same or even a bit larger number of layers, stacking multiple residual network units cannot beat the current one-unit structure, which indicates that the uninterrupted layer structure is more efficient to extract the high-order features in this case.

\emph{3) Stage 3:} Denote $\mathcal{B}$ as the index set of the inactive RIS elements with $N_{2}=|\mathcal{B}|$. The aim of this stage is to infer the equivalent cascaded channel associated with $\mathcal{B}$, $\mathbf{G}_{\mathcal{B}}$, from $\hat{\mathbf{G}}_{\mathcal{A},\textrm{dnn}}$. A straightforward way to carry out this task is inputting $\hat{\mathbf{G}}_{\mathcal{A},\textrm{dnn}}$ into a CNN to approximate $\mathbf{G}_{\mathcal{B}}$. However, it is challenging to uncover this matrix mapping relationship since the structure of the equivalent cascaded channel is quite different from that of the single-hop channel. This inspires us to dissect the channel structure and find the more suitable input that the neural network can digest well.

Since $\mathbf{G}=\mathbf{H}_{\textrm{RB}}\textrm{diag}(\mathbf{h}_{\textrm{UR}})$, we start with analyzing the structure of $\mathbf{H}_{\textrm{RB}}$. According to Saleh-Valenzuela (SV) model, $\mathbf{H}_{\textrm{RB}}$ is given by
\vspace{-0.1cm}
\begin{eqnarray}
\label{eqn_H_RB}
\mathbf{H}_{\textrm{RB}}=\sqrt{\frac{MN}{L_{\textrm{RB}}}}\sum_{l=1}^{L_{\textrm{RB}}}\alpha_{\textrm{RB},l}\mathbf{a}_{\textrm{B}}(\theta_{l})\mathbf{a}^{H}_{\textrm{R}}(\varphi_{l}),
\end{eqnarray}
where $L_{\textrm{RB}}$, $\alpha_{\textrm{RB},l}$, $\theta_{l}$, and $\varphi_{l}$ denote number of main paths, the complex gain of the $l$th path, the azimuth angles of arrival and departure (AoA/AoD) at the BS and the RIS, respectively. The response vector $\mathbf{a}_{\textrm{B}}(\theta_{l})$ can be further expressed as $\mathbf{a}_{\textrm{B}}(\theta_{l})= \frac{1}{\sqrt{M}}\bigl[1,e^{-j2\pi\frac{d}{\lambda}\sin(\theta_l)},\ldots,e^{-j2\pi\frac{d}{\lambda}(M-1)\sin(\theta_l)}\bigr]^{T}$ with $d$ and $\lambda$ denoting the space between the adjacent antennas at the BS and the wavelength of the carrier frequency, respectively. Then the $m$th row of $\mathbf{H}_{\textrm{RB}}$ can be written as
\vspace{-0.1cm}
\begin{eqnarray}
\label{eqn_h_RBm}
\mathbf{h}_{\textrm{RB},m}=\sqrt{\frac{N}{L_{\textrm{RB}}}} \sum_{l=1}^{L_{\textrm{RB}}}\alpha_{\textrm{RB},l}e^{-j2\pi\frac{d}{\lambda}(m-1)\sin(\theta_l)}\mathbf{a}^{H}_{\textrm{R}}(\varphi_{l}).
\end{eqnarray}
It is obvious that each row of $\mathbf{H}_{\textrm{RB}}$ exhibits a unified form containing all the channel information between the RIS and the BS. Consequently, the $m$th row of $\mathbf{G}$, which is given by $\mathbf{g}_{m}=\mathbf{h}_{\textrm{RB},m}\textrm{diag}(\mathbf{h}_{\textrm{UR}})$, $\forall m\in\{1,\ldots,M\}$, inherits this property with all the channel information of the RIS-relayed channel included.

Denote $\mathbf{g}_{\mathcal{A},m}$ and $\mathbf{g}_{\mathcal{B},m}$ as the parts of $\mathbf{g}_{m}$ corresponding to $\mathcal{A}$ and $\mathcal{B}$, respectively. Based on the perspective view of the channel mentioned above, we can focus on $\mathbf{g}_{\mathcal{A},m}$ and $\mathbf{g}_{\mathcal{B},m}$, instead of $\mathbf{G}_{\mathcal{A}}$ and $\mathbf{G}_{\mathcal{B}}$, to extract the inherence underlaying the mapping relationship, which facilitates the DNN design and training with significantly improved prediction accuracy. Specifically, as shown in Fig.~\ref{CSI_framework}, we design an \textbf{I}nactive \textbf{R}IS Channel \textbf{P}rediction DNN (IRP-DNN) to approximate $\mathbf{g}_{\mathcal{B},m}$ by using $\hat{\mathbf{g}}_{\mathcal{A},m,\textrm{dnn}}$, i.e., the $m$th row of $\hat{\mathbf{G}}_{\mathcal{A},\textrm{dnn}}$, as the input with $m$ randomly selected from $\{1,\ldots,M\}$. The loss function for IRP-DNN training is expressed as
\vspace{-0.1cm}
\begin{eqnarray}
\label{eqn_mse_IRP}
\mathcal{L}_{\textrm{IRP}}=\frac{1}{N_{\textrm{tr}}}\sum_{n=1}^{N_{\textrm{tr}}} \|\mathbf{g}_{\mathcal{B},m}^{(n)}-\hat{\mathbf{g}}_{\mathcal{B},m,\textrm{dnn}}^{(n)}\|^2,
\end{eqnarray}
where $\hat{\mathbf{g}}_{\mathcal{B},m,\textrm{dnn}}^{(n)}$ denotes the output of IRP-DNN. Since we focus on the level of vector mapping instead of matrix mapping, IRP-DNN is designed in a FC structure. There are three hidden layers applying ReLU activation function and BN while no activation function is applied in the output layer.

\vspace{-0.1cm}
\subsection{Online Testing}

After offline training, the CSI acquisition framework will be deployed for online testing. From Fig.~\ref{CSI_framework}, $\hat{\mathbf{h}}_{\textrm{UB,ls}}$, which is obtained from $\mathbf{y}_{1}$, is first refined by DE-DNN to output $\hat{\mathbf{h}}_{\textrm{UB,dnn}}$, based on which the component of the direct channel will be stripped from $\mathbf{Y}_{\mathcal{A}}$. Then $\hat{\mathbf{G}}_{\mathcal{A},\textrm{ls}}$ is obtained and will be further purified via ARE-DNN to yield $\hat{\mathbf{G}}_{\mathcal{A},\textrm{dnn}}$. Afterward, each row of $\hat{\mathbf{G}}_{\mathcal{A},\textrm{dnn}}$ is input into IRP-DNN sequentially to predict the corresponding $\hat{\mathbf{g}}_{\mathcal{B},m,\textrm{dnn}}$. Vertically stacking $\hat{\mathbf{g}}_{\mathcal{B},m,\textrm{dnn}}$ ($\forall m\in\{1,\ldots,M\}$) constructs $\hat{\mathbf{G}}_{\mathcal{B},\textrm{dnn}}$ and combining $\hat{\mathbf{G}}_{\mathcal{A},\textrm{dnn}}$ and $\hat{\mathbf{G}}_{\mathcal{B},\textrm{dnn}}$ finally obtains $\hat{\mathbf{G}}_{\textrm{dnn}}$. It can be seen that the proposed CSI acquisition framework is able to estimate $\mathbf{h}_{\textrm{UB}}$ and $\mathbf{G}$ with high accuracy while the pilot overhead can be reduced by the ratio of $\frac{N_1+1}{N+1}$.

In addition, the computational complexities of DE-DNN, ARE-DNN, and IRP-DNN are $\mathcal{O}(\sum_{i=2}^{L_{\textrm{DE}}}N^{\textrm{DE}}_{i-1}N^{\textrm{DE}}_{i})$, $\mathcal{O}(MN_1^2\\+MN_1\sum_{i=1}^{L_{\textrm{ARE}}}K_{i}^2F_{i-1}F_{i})$, and $\mathcal{O}(M\sum_{i=2}^{L_{\textrm{IRP}}}N^{\textrm{IRP}}_{i-1}N^{\textrm{IRP}}_{i})$, respectively, where $L_{\textrm{DE}}$, $L_{\textrm{ARE}}$, and $L_{\textrm{IRP}}$ denote the numbers of (convolutional) layers of DE-DNN, ARE-DNN, and IRP-DNN, $N^{\textrm{DE}}_{i}$ and $N^{\textrm{IRP}}_{i}$ denotes the corresponding numbers of neurons of the $i$th layer, $K_{i}$ is the side length of the filters used by the $i$th convolutional layer, $F_{i-1}$ and $F_{i}$ denote the numbers of input and output feature maps of the $i$th convolutional layer.

\vspace{-0.1cm}
\section{Simulation Results}

\begin{table}[!t]
\centering
\caption{Structures of DE-DNN, ARE-DNN, and IRP-DNN}
\label{table_1}
\begin{tabular}{p{0.55cm}<{\centering}|c|c|c|c}
\hline
~ & \makecell{Layer\\type} & \makecell{Tensor size} & \makecell{Kernel\\size} & \makecell{Activation \\function}\\
\hline
\multirow{5}{*}{\makecell{DE-\\DNN}} & Input & $2M$ & - & - \\
\cline{2-5}
~ & Dense & 64 & - & ReLU\\
\cline{2-5}
~ & Dense & 128 & - & ReLU\\
\cline{2-5}
~ & Dense & 64 & - & ReLU\\
\cline{2-5}
~ & Output & $2M$ & - & -\\
\hline
\multirow{3}{*}{\makecell{ARE-\\DNN}} & Input & $M\times N_1\times2$ & - & - \\
\cline{2-5}
~ & \makecell{ZP Conv.\\($7$ layers)} & $M\times N_1\times64$ & $3\times 3$ & ReLU\\
\cline{2-5}
~ & Output & $M\times N_1\times2$ & $3\times 3$ & -\\
\hline
\multirow{5}{*}{\makecell{IRP-\\DNN}} & Input & $2N_1$ & - & - \\
\cline{2-5}
~ & Dense & 128 & - & ReLU\\
\cline{2-5}
~ & Dense & 256 & - & ReLU\\
\cline{2-5}
~ & Dense & 256 & - & ReLU\\
\cline{2-5}
~ & Output & $2N_2$ & - & - \\
\hline
\end{tabular}
\vspace{-0.4cm}
\end{table}

\begin{figure*}[htbp]
\centering
\begin{minipage}[t]{2.3in}
\setlength{\abovecaptionskip}{-0.1cm}
\includegraphics[width=2.3in]{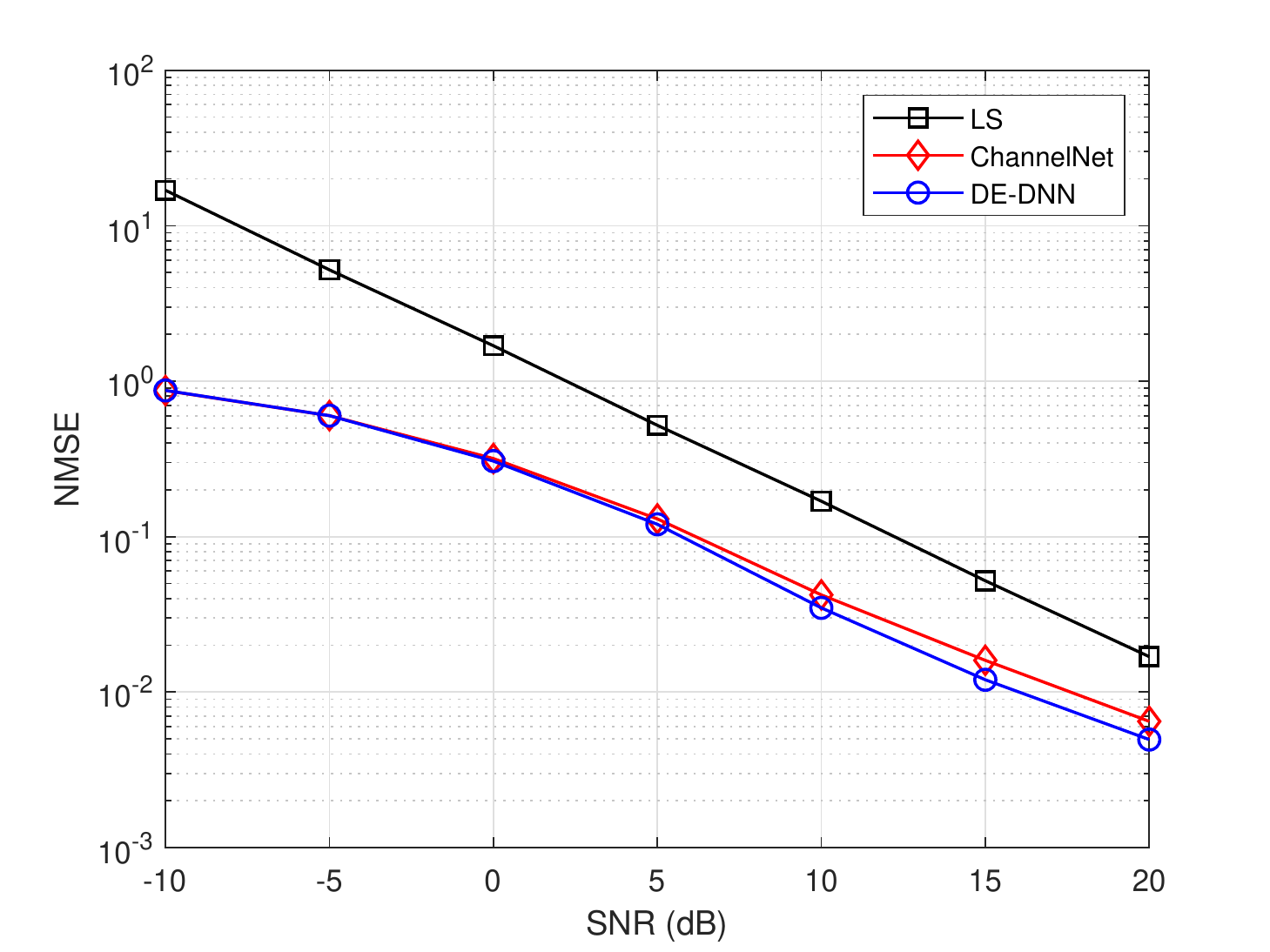}
\caption{NMSE versus SNR for the direct channel $\mathbf{h}_{\textrm{UB}}$.}
\label{nmse_hUB}
\end{minipage}
\hspace{0.3ex}
\begin{minipage}[t]{2.3in}
\setlength{\abovecaptionskip}{-0.1cm}
\includegraphics[width=2.3in]{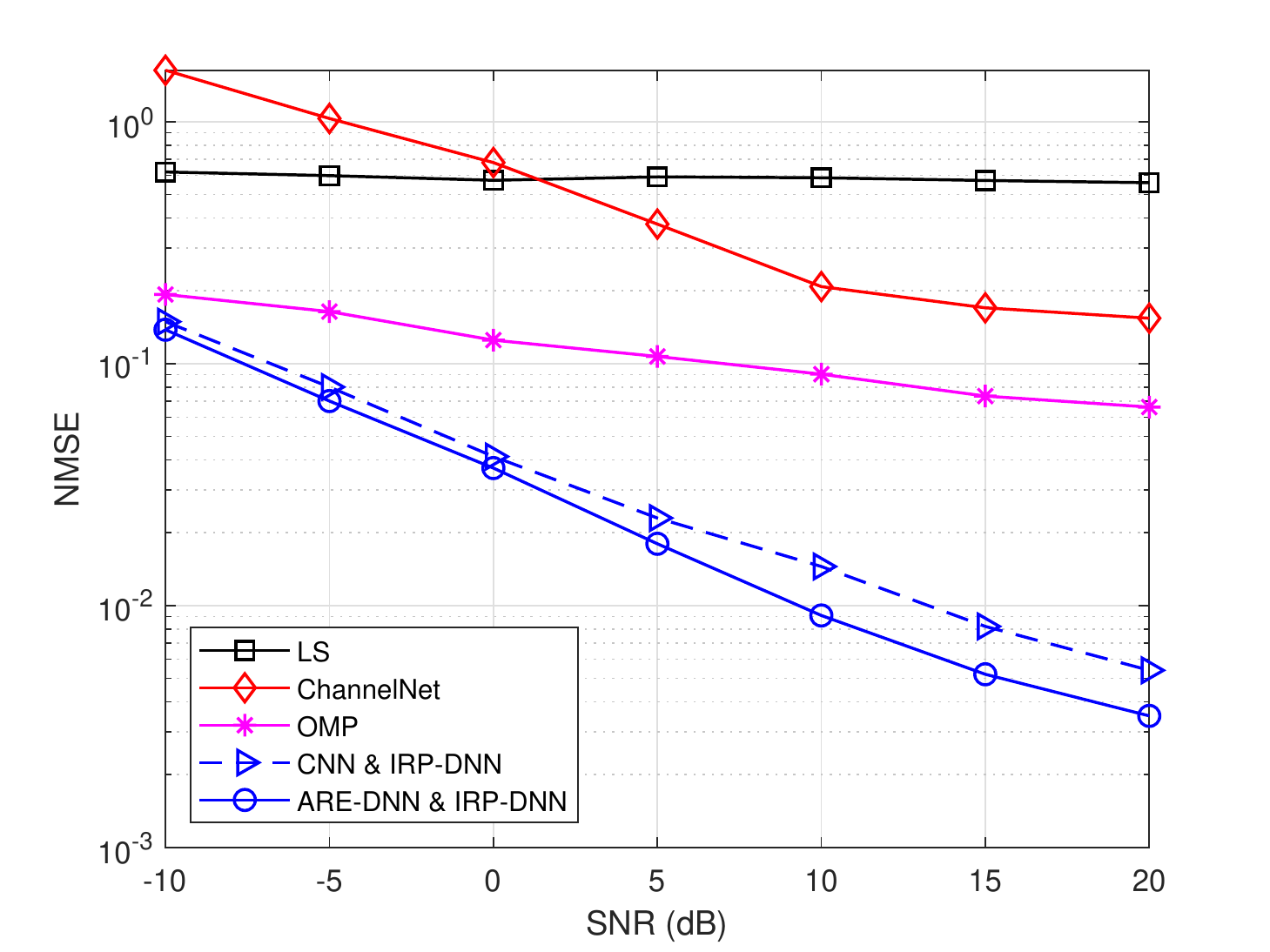}
\caption{NMSE versus SNR for the equivalent cascaded channel $\mathbf{G}$.}
\label{nmse_Gc}
\end{minipage}
\hspace{0.3ex}
\begin{minipage}[t]{2.3in}
\setlength{\abovecaptionskip}{-0.1cm}
\includegraphics[width=2.3in]{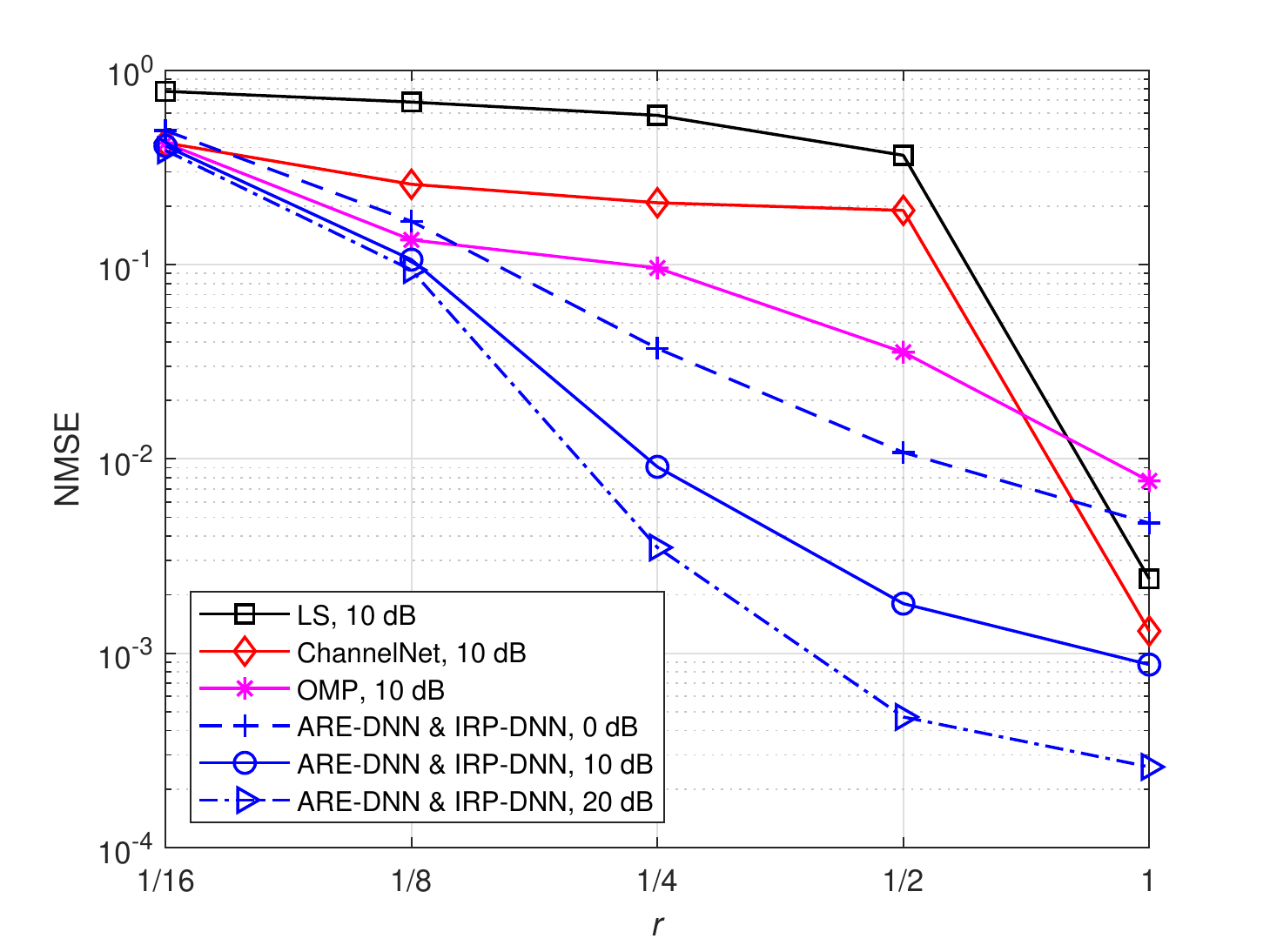}
\caption{NMSE versus the pilot overhead ratio for $\mathbf{G}$.}
\label{nmse_Gc_vs_r}
\end{minipage}
\hspace{0.5ex}
\vspace{-0.2cm}
\end{figure*}

In this section, numerical results are presented to validate the proposed DL-based CSI acquisition framework. The baseline schemes for comparison include the LS estimator \cite{B. Zheng}, \cite{T. L. Jensen}, orthogonal matching pursuit (OMP) \cite{P. Wang} and ChannelNet \cite{A. M. Elbir}. The numbers of BS antennas and RIS elements are set as $M=16$ and $N=128$, respectively. The numbers of paths of $\mathbf{h}_{\textrm{UB}}$, $\mathbf{h}_{\textrm{UR}}$, and $\mathbf{H}_{\textrm{RB}}$ are set as $3$. The average noise power, $\sigma_{0}^2$, is normalized to $1$ and the transmit power, $P$, is set as a relative value with respect to $\sigma_{0}^2$. The signal-to-noise ratio (SNR) in Figs.~\ref{nmse_hUB} and \ref{nmse_Gc} is defined as $\textrm{SNR}=10\lg\frac{P}{\sigma_{0}^2}$ (dB). Without loss of generality, assume that the pathloss has been absorbed into the transmit power for simplicity. Then the average power gain of each path of $\mathbf{h}_{\textrm{UB}}$, $\mathbf{h}_{\textrm{UR}}$, and $\mathbf{H}_{\textrm{RB}}$ is set as $1$ \cite{Z. He}. For the proposed framework, the training, validation, and testing sets contain $90,000$, $10,000$, and $10,000$ samples, respectively. Adam is applied as the optimizer and the batch size is set as $128$. The training of each DNN therein lasts $300$ epochs with the initial learning rates $1\times10^{-3}$ and $1\times10^{-4}$ for the first $200$ epochs and the remaining $100$ epochs, respectively. The detailed DNN structures are listed in Table~\ref{table_1}.
For offline training, the label of DE-DNN, $\mathbf{h}_{\textrm{UB}}$, is generated according to the SV channel model and the input, $\hat{\mathbf{h}}_{\textrm{UB,ls}}$, is generated by $\hat{\mathbf{h}}_{\textrm{UB,ls}}=\frac{\mathbf{y}_{1}}{\sqrt{P}}$. Then the label of ARE-DNN, $\mathbf{G}_{\mathcal{A}}$, is obtained by fetching the corresponding columns from $\mathbf{G}$ as per the indexes in $\mathcal{A}$. The input, $\hat{\mathbf{G}}_{\mathcal{A},\textrm{ls}}$, is calculated as (\ref{eqn_GA_ls}), where $\mathbf{h}_{\textrm{UB}}$ is the training label of DE-DNN and $\hat{\mathbf{h}}_{\textrm{UB,dnn}}$ is the corresponding output. The label of IRP-DNN, $\mathbf{g}_{\mathcal{B},m}$, is the $m$th row of $\mathbf{G}_{\mathcal{B}}$, where $\mathbf{G}_{\mathcal{B}}$ is the complementary matrix of the training label, $\mathbf{G}_{\mathcal{A}}$, and $m$ is arbitrarily selected from $\{1,\ldots,M\}$. The input, $\hat{\mathbf{g}}_{\mathcal{A},m,\textrm{dnn}}$, is the $m$th row of $\hat{\mathbf{G}}_{\mathcal{A},\textrm{dnn}}$, which is the output of ARE-DNN when the input is $\hat{\mathbf{G}}_{\mathcal{A},\textrm{ls}}$. The testing samples are generated similarly to training samples, but different path gains and AoAs/AoDs in the SV model are used.
The normalized mean-squared error (NMSE) is used to evaluate the CSI acquisition performance and can be expressed as $\textsf{NMSE}=\mathbb{E}\Bigl\{\frac{\|\mathbf{h}_{\textrm{UB}}-\hat{\mathbf{h}}_{\textrm{UB,dnn}}\|^2}{\|\mathbf{h}_{\textrm{UB}}\|^2}\Bigr\}$ for $\mathbf{h}_{\textrm{UB}}$ and $\textsf{NMSE}=\mathbb{E}\Bigl\{\frac{\|\mathbf{G}-\hat{\mathbf{G}}_{\textrm{dnn}}\|_F^2}{\|\mathbf{G}\|_F^2}\Bigr\}$ for $\mathbf{G}$. The pilot overhead ratio for estimating $\mathbf{G}$ is defined as $r=\frac{N_1}{N}$.

Fig.~\ref{nmse_hUB} plots the NMSE performance versus SNR for $\mathbf{h}_{\textrm{UB}}$. From Fig.~\ref{nmse_hUB}, both the ChannelNet and the proposed DE-DNN are effective to refine the LS estimate and further improve the accuracy. Compared with the ChannelNet, DE-DNN achieves almost same or even better performance, indicating that the DNN with FC structure is adequate to estimate the direct channel with high accuracy.

The estimation performance of $\mathbf{G}$ versus SNR is shown in Fig.~\ref{nmse_Gc} with $r=\frac{1}{4}$. The LS estimator keeps error floor over the considered SNR regime because the performance will be poor regardless of the SNR level once the pilot overhead $N_1$ is less than $N$. The ChannelNet is based on the LS estimation and thus also performs unsatisfactorily. The OMP approach outperforms the LS estimator and ChannelNet but cannot provide very accurate CSI due to the complication of the cascaded RIS channel. By contrast, the proposed framework divides the acquisition of $\mathbf{G}$ into estimation and prediction stages with respective dedicated DNNs, i.e., ARE-DNN and IRP-DNN, which can learn the features of the RIS-relayed channel more comprehensively and thus achieve the superior performance. In addition, the performance will be a little worse if the residual network structure of ARE-DNN is replaced by a CNN with the same number of layers, which indicates the effectiveness of the residual network structure.

In Fig.~\ref{nmse_Gc_vs_r}, we further investigate the NMSE performance of $\mathbf{G}$ versus the pilot overhead ratio, $r$. The proposed framework (ARE-DNN and IRP-DNN) always outperforms the LS estimator, ChannelNet, and OMP at different values of $r$, which demonstrates that the proposed framework is more robust to the reduction of the pilot overhead. For the proposed framework, the performance gap between different SNRs increases with $r$ since the dominating factor turns into SNR from the pilot overhead.

Considering the time complexity for practical implementation, the runtimes of the proposed scheme and the ChannelNet are $1.37\times10^{-4}$ seconds and $2.75\times10^{-4}$ seconds, respectively, on the GTX 2080Ti GPU  while the runtime of the OMP is $5.87\times10^{-3}$ seconds on the Intel(R) Core(TM) i7-3770 CPU. The proposed scheme consumes minimum runtime owing to elaborated design and efficient parallel computing.

\vspace{-0.2cm}
\section{Conclusion}

In this article, we develop a three-stage CSI acquisition framework for the RIS-aided MIMO uplink system based on an elaborated synthetic DNN. It includes three dedicated DNNs in charge of estimating the direct channel, estimating the cascaded channel for active RIS elements, and predicting the cascaded channel for inactive RIS elements, respectively. The three DNNs with specialized structures are trained and hook up sequentially. Simulation results show that the proposed CSI acquisition framework can achieve superior performance without relying on high pilot overhead and exact knowledge on channel statistics. In future work, the proposed framework can be extended to the broadband channel exploiting frequency correlation or common sparsity \cite{P. Dong}, \cite{S. Liu}.



\ifCLASSOPTIONcaptionsoff
  \newpage
\fi

\end{document}